\documentclass[letterpaper,aps,superscriptaddress,showpacs,floatfix,twocolumn,10pt]{revtex4-1}
\usepackage{graphicx}
\usepackage{amsmath}
\usepackage{graphicx}
\usepackage{float}

\begin{document}
\title{Constraining the Eq. of State of Super-Hadronic Matter from Heavy-Ion Collisions}
\author{Scott Pratt}
\author{Evan Sangaline}
\affiliation{Department of Physics and Astronomy and National Superconducting Cyclotron Laboratory\\
Michigan State University, East Lansing, MI 48824, USA}
\author{Paul Sorensen}
\author{Hui Wang}
\affiliation{Brookhaven National Laboratory, Upton, New York 11973, USA}
\date{\today}


\begin{abstract}
The equation of state of QCD matter for temperatures near and above the quark-hadron transition ($\sim 165$ MeV) is inferred within a Bayesian framework through the comparison of data from the Relativistic Heavy Ion Collider and from the Large Hadron Collider to theoretical models. State-of-the-art statistical techniques are applied to simultaneously analyze multiple classes of observables while varying 14 independent model parameters. The resulting posterior distribution over possible equations of state is consistent with results from lattice gauge theory.
\end{abstract}

\maketitle

\section{Introduction}

Relativistic heavy ion collisions have been proposed as a means for investigating the equation of state of hot matter. For fixed target energies of $E/A\lesssim 10$ GeV, analyses of heavy ion collisions have significantly constrained the compressibility of 
dense hadronic matter \cite{Danielewicz:2002pu} for temperatures $\lesssim 100$ MeV. Higher energy collisions probe conditions near and above the transition temperature, where lattice calculations have shown that in a narrow temperature band, $150<T<200$ MeV, the scalar quark condensate melts \cite{Bazavov:2011nk}, the degrees of freedom change \cite{Ratti:2011au}, and the speed of sound has a minimum \cite{Bazavov:2014pvz}. In fact, for some time the transition was postulated to contain a first-order phase transition accompanied by a sizable latent heat. 

In contrast to the progress of lattice calculations, experimental determination of the equation of state at high temperature has remained semi-quantitative. The stunted progress has not been due to a shortage of experimental observables that are known to be sensitive to the equation of state. Van Hove associated the dependence of the mean transverse momentum, $\langle p_t\rangle$, as a function of multiplicity as tool for determining the equation of state \cite{VanHove:1982vk}. Two-particle femtoscopic correlations were proposed as a signal for a first-order phase transition \cite{Pratt:1986cc}. Measurements of azimuthal elliptic flow, which are now mainly associated with determining the viscosity, were also shown to be sensitive to the equation of state \cite{Sorge:1996pc,Sorge:1998mk}. Multiplicities, which are related to entropy, have also been used to constrain the equation of state \cite{Pal:2003rz}. Although femtoscopic analyses have shown that a first order equation of state with a large latent heat is highly unlikely \cite{Lisa:2005dd}, and that an extremely stiff equation of state, such as that of a pion gas, is also inconsistent with data \cite{Pratt:2008qv}, a more quantitative statement of how well the equation of state is constrained has proven elusive. Even if analysis of experimental data cannot compete with lattice calculations in determining the equation of state for perfectly equilibrated matter, constraining the equation of state by experiment can help validate the statement that the matter created in heavy-ion collisions behaves like an equilibrated quark gluon plasma.

The road block to turning these sensitivities into a more robust and rigorous determination of the equation of state has been the intertwined dependencies between the many unknown features and parameters of the model, and the numerous classes of measurement. Two developments now make this next step possible. First, the models used to describe the bulk behavior have converged to a standard framework based on relativistic viscous hydrodynamics for the evolution of the high temperature region, $\gtrsim 165$ MeV, \cite{Heinz:2013wva} coupled to a microscopic simulation of the lower temperature hadronic stage based on binary collisions. The initial evolution, which feeds into the hydrodynamic description, remains rather undefined, but one can represent those uncertainties parametrically. The second development is in the statistical methodologies and tools required to compare heterogenous data to models where a large number of parameters are required to encapsulate the many model uncertainties. Here we use the statistical tools described in \cite{Novak:2013bqa} to constrain 14 parameters via a Markov-chain Monte Carlo. The statistical tools are based on a Gaussian-process model emulator, which allows one to estimate observables for a given point in parameter space by interpolating from a fixed number of full-model runs. 

\section{Methodology}

Here we report on comparisons of model calculations to data from Au+Au collisions from the highest RHIC (Relativistic Heavy Ion Collider) energy, 100$A$ GeV + 100$A$ GeV, and from Pb+Pb collisions at the Large Hadron Collider (LHC), 1.38$A$ TeV + 1.38$A$ TeV. The hydrodynamic and hadronic simulations were the same as those used in \cite{Novak:2013bqa} to analyze RHIC data. The analysis involves 14 parameters, two of which vary the equation of state. The statistical method returns a sampling of the 14-dimensional space that is weighted by the likelihood,
\begin{equation}
\label{eq:likelihood}
{\cal L}(\vec{x})\sim \prod_i \exp\left\{
-\left((z_i^{\rm(mod)}(\vec{x})-z_i^{\rm(exp)}\right)^2/2.
\right\}.
\end{equation}
Here, $\vec{x}$ is the 14-dimensional vector describing a point in parameter space and $z_i$ are principal components of the observables where each observable $y_i$ is first scaled by $\sigma_i$ which describes the uncertainty one assigns to the comparison of the model to experiment, with $\sigma_i$ accounting for both experimental uncertainties and the error one might associate with the model missing some of the physics. Here, the uncertainties were all chosen to be 6\% of each observable. Changing this to 9\% only modestly affected the final result. The largest source of uncertainty derives from the unknown impact of missing physics. These shortcomings will be discussed further below.

Constraining the equation of state is the principal goal of this study. The equation of state was chosen to be consistent with that of a hadron gas for a temperature of 165 MeV, which is the temperature at which the hydrodynamic description switched to the microscopic hadronic simulation. At the high-energy densities considered here, one can neglect any small excess of baryons to antibaryons and the equation of state can be expressed in terms of a single variable such as the energy density $\epsilon$. For temperatures above 165 MeV, the speed of sound squared was parameterized to allow for a large range of equations of state,
\begin{eqnarray}
\label{eq:EoSform}
c_s^2(\epsilon)&=&c_s^2(\epsilon_h)+\left(\frac{1}{3}-c_s^2(\epsilon_h)\right)\frac{X_0x+x^2}{X_0x+x^2+X'^2},\\
\nonumber
X_0&=&X'Rc_s(\epsilon)\sqrt{12},~~x\equiv\ln\epsilon/\epsilon_h,
\end{eqnarray}
where $\epsilon_h$ is the energy density corresponding to $T=165$ MeV. The two parameters $R$ and $X'$ describe the behavior of the speed of sound at energy densities above $\epsilon_h$. Whereas $R$ describes how the speed of sound rises or falls for small $x$, $X'$ describes how quickly the speed of sound eventually approaches $1/3$ at high temperature. Once given $c_s^2(\epsilon)$, thermodynamic relations provide all other representations of the equation of state. Runs were performed for $0.5<X'<5$, and with $-0.9<R<2$. In the limit $R\rightarrow -1$ the speed of sound will have a minimum of zero.

Ten of the 14 model parameters described the initial stress-energy tensor and flow used to describe the initial state and instantiate the hydrodynamic calculation, with 5 separate parameters describing the initial state for each beam energy.  Three parameters varied the transverse profile of the initial energy density at each beam energy: a weight between two saturation pictures, a normalization for the initial energy density, and a screening parameter. These three parameters, along with a parameter used to vary the initial flow, are described in \cite{Novak:2013bqa}. The fifth parameter describes the initial anisotropy of the stress energy tensor and was varied so that the longitudinal pressure, $T_{zz}$, could vary between zero and the pressure $P$. The viscosity at $T=165$ MeV and its temperature dependence were described by two parameters as was done in \cite{Novak:2013bqa} and the final two parameters varied the equation of state. 

The details of both the physical model and the statistical method are described in \cite{Novak:2013bqa}. The calculations shown here were based on 1200 full-model runs. Thirty observables, 15 for RHIC data and 15 for the LHC, were related to spectra, elliptic flow and femtoscopic source sizes. Observables were calculated for two centralities, 20-30\% and 0-5\% for both the RHIC and LHC cases. At each centrality the spectral observables were the mean transverse momenta, $\langle p_t\rangle$, for pions, kaons and protons, and the yield for pions. The three femtoscopic sizes, averaged over the experimentally analyzed momentum range, $R_{\rm out}$, $R_{\rm side}$ and $R_{\rm long}$ described the dimensions of the outgoing phase space cloud of particles with the same momenta. The
$\langle p_t\rangle$-weighted measurement of the elliptic flow, $v_2=\langle \cos 2\phi\rangle$ quantified the preference for emitting particles in the reaction plane $(\phi=0~{\rm or}~180^\circ)$. Because the model used initial energy profiles that were smoothed by considering the averaged positions of incoming nucleons within a nucleus, rather than more realistic lumpy, or fluctuating, initial conditions, the model had to scale up its predictions for elliptic flow by a factor of 1.10. This accounts for the fact that the fluctuations result in larger initial transverse elliptic asymmetries which then lead proportionally to larger flows. The correction factor was quantitatively evaluated assuming a linear response in $v_2$ to initial eccentricity and found to be minimized in the 20-30\% centrality class. The $v_2$ analysis was confined to the 20-30\% centrality to minimize the effect of fluctuating initial conditions.

The first 1000 runs were chosen semi-randomly throughout the 14-dimensional parameter space according to latin hyper-cube sampling. The thirty observables were then reduced to 14 principal components, which captured over 99.9\% of the variance. Identically to what was done in \cite{Novak:2013bqa}, these principal components were interpolated from the 1000 runs using a Gaussian process emulator during a Markov chain Monte Carlo (MCMC) exploration of the parameter space. This yielded a posterior sampling of the parameter space, i.e. a sampling that was weighted by the likelihood to reproduce the measured observables. A sampling of 50 points in parameter space was then chosen according to the posterior distribution and evaluated with the full model. Real model values were then compared to the emulated values at these 50 points to validate the emulator in the regions of high likelihood which are most important in correctly determining the posterior distribution. The emulator was then retrained using the 1050 runs and the validation procedure was repeated three additional times resulting in a total of 1200 model runs. The emulator's accuracy in each case was found to be a few tenths of one unit when determining $\ln(\cal{L})$ in Eq. (\ref{eq:likelihood}). The results shown here use emulation based upon the full model runs at these 1200 points in parameter space, 200 of which are distributed according to the posterior distribution.

\begin{figure}
\includegraphics[width=0.4\textwidth]{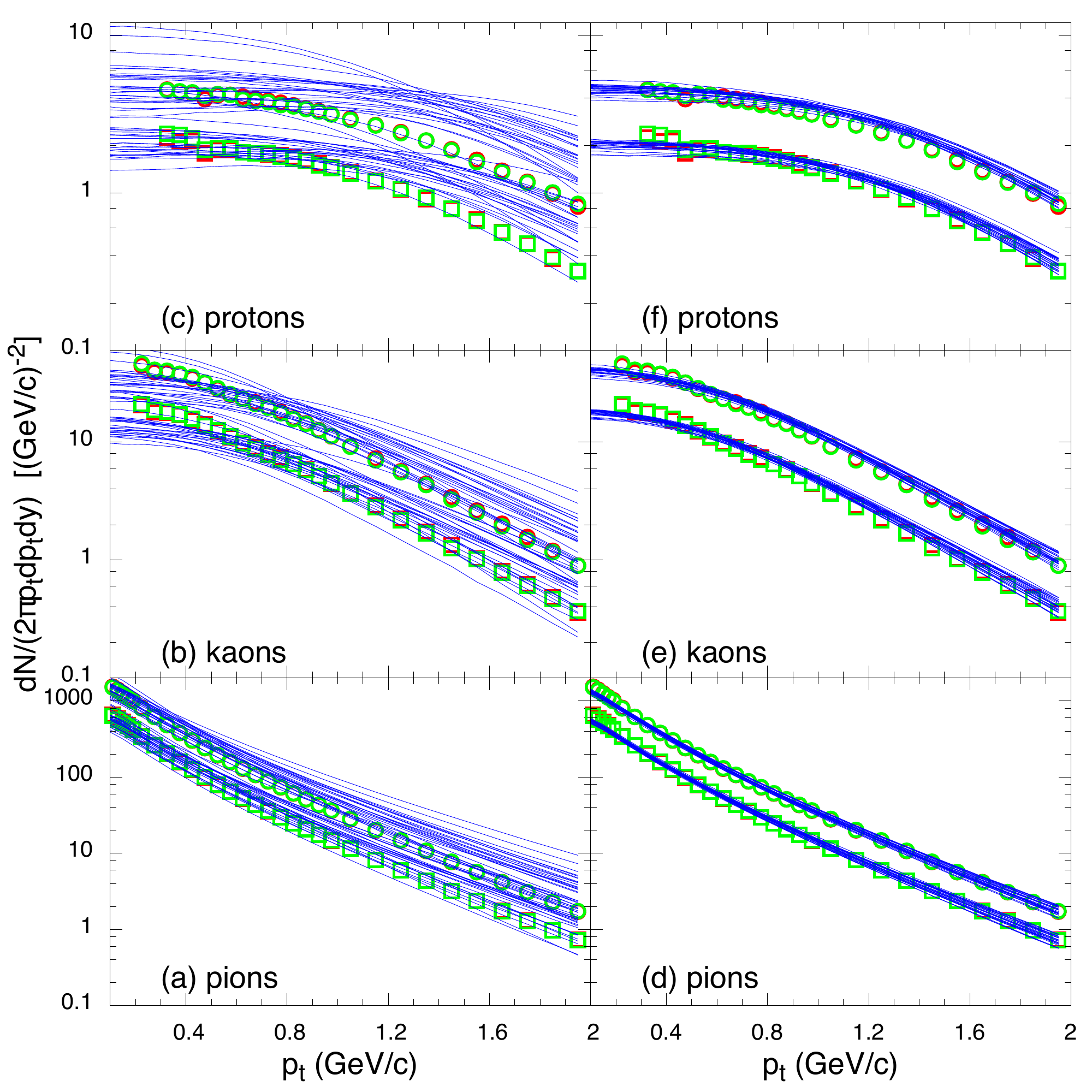}
\caption{\label{fig:spectra_lhc}
Twenty pion, kaon and proton spectra as measured by the ALICE collaboration at the LHC (circles for 0-5\% and squares for 20-30\%) \cite{Abelev:2013vea} are compared to model predictions using parameters randomly taken from the prior parameter space (panels a-c) and using parameters weighted by the likelihood (d-f).
}
\end{figure}
\begin{figure}
\includegraphics[width=0.45\textwidth]{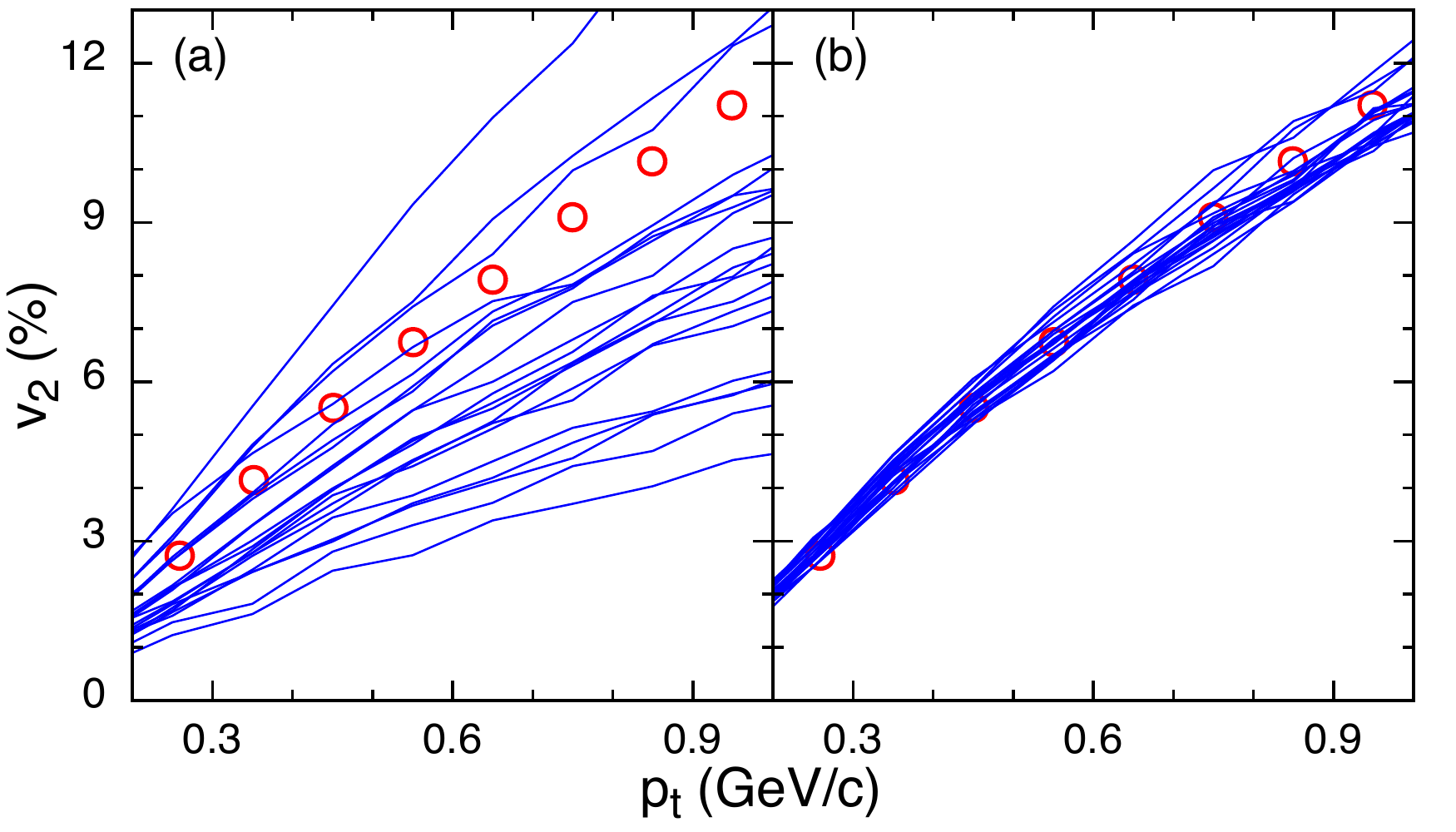}
\caption{\label{fig:v2_lhc}
The pion azimuthal anisotropy $v_2$, often referred to as elliptic flow, from ALICE \cite{Aamodt:2010pa} for the 20-30\% centrality (circles) are compared to model predictions using parameters randomly taken from the prior parameter space (a), and weighted by the likelihood (b).
}
\end{figure}
\begin{figure}
\includegraphics[width=0.4\textwidth]{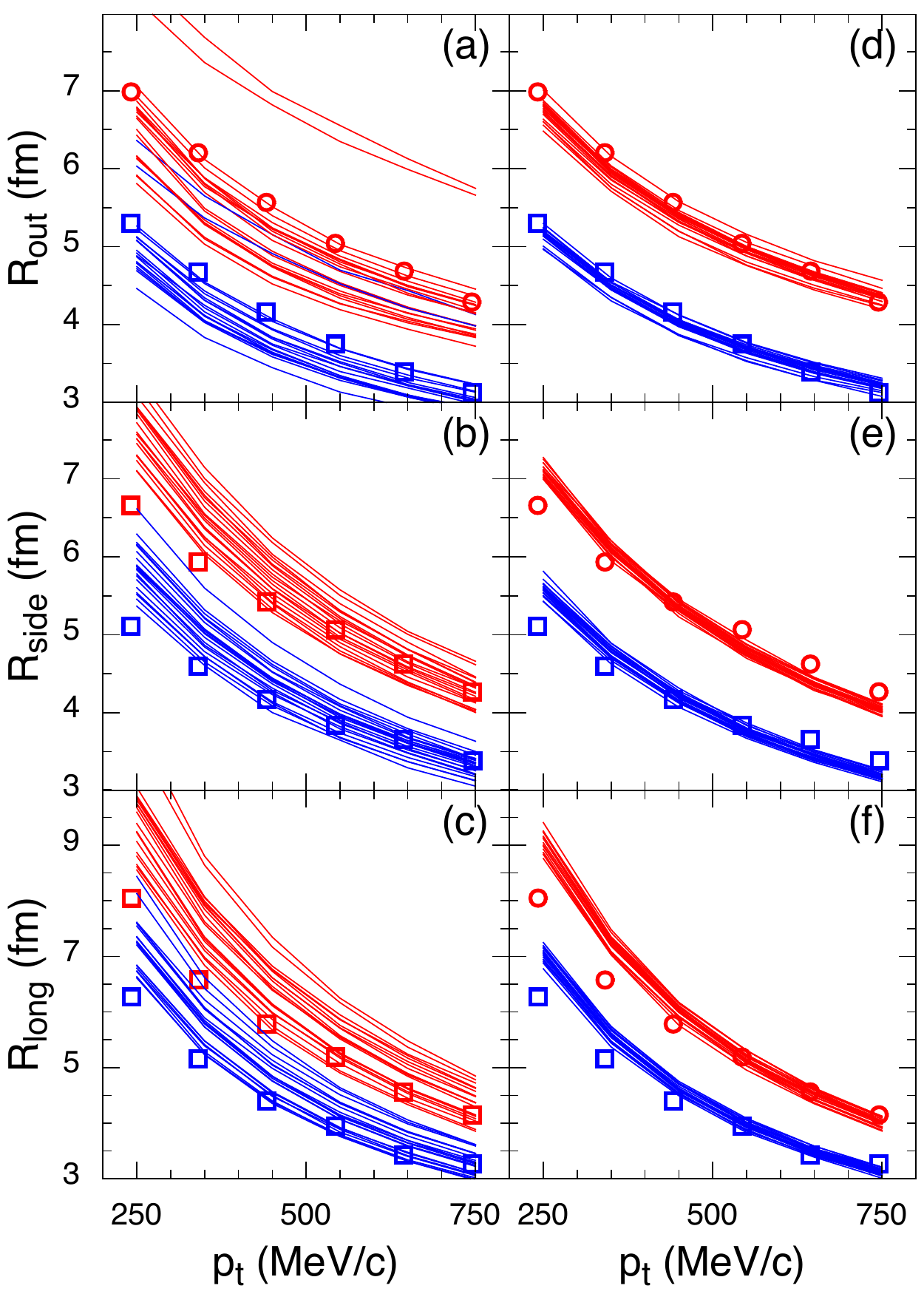}
\caption{\label{fig:hbt_lhc}
Two-particle femtoscopic source sizes from ALICE \cite{Graczykowski:2014hoa} (circles for 0-5\% and squares for 20-30\% centrality) are compared to model predictions using parameters randomly taken from the prior parameter space (a-c), and weighted by the likelihood (d-f).}
\end{figure}

\section{Results}

The ability of the procedure to accurately identify likely regions of parameter space is illustrated in Figs. \ref{fig:spectra_lhc}, \ref{fig:v2_lhc} and \ref{fig:hbt_lhc} by comparing both full model calculations at 20 random points in parameter space and then again at 20 points chosen proportional to the likelihood defined in Eq. (\ref{eq:likelihood}). Calculations are compared to  ALICE Collaboration at the LHC. Similarly good representations of the experimental data are found for RHIC data, with results very similar to those in \cite{Novak:2013bqa}. The procedure readily identified regions of parameter space that matched the experimental data within the 6\% uncertainty assumed here. Nonetheless, it appears that the procedure finds spectra that have transverse momenta that are a few percent higher than the the experiment, and femtoscopic source sizes that are a few percent larger. This suggests the femtoscopic data and the spectra are competing for agreement with the model, as slightly more explosive models would better match the femtoscopic observations, while less explosive models would better reproduce the spectra. This implies that improved physics might be needed if one were to reproduce the experimental results much better than 6\%.

\begin{figure}
\includegraphics[width=0.4\textwidth]{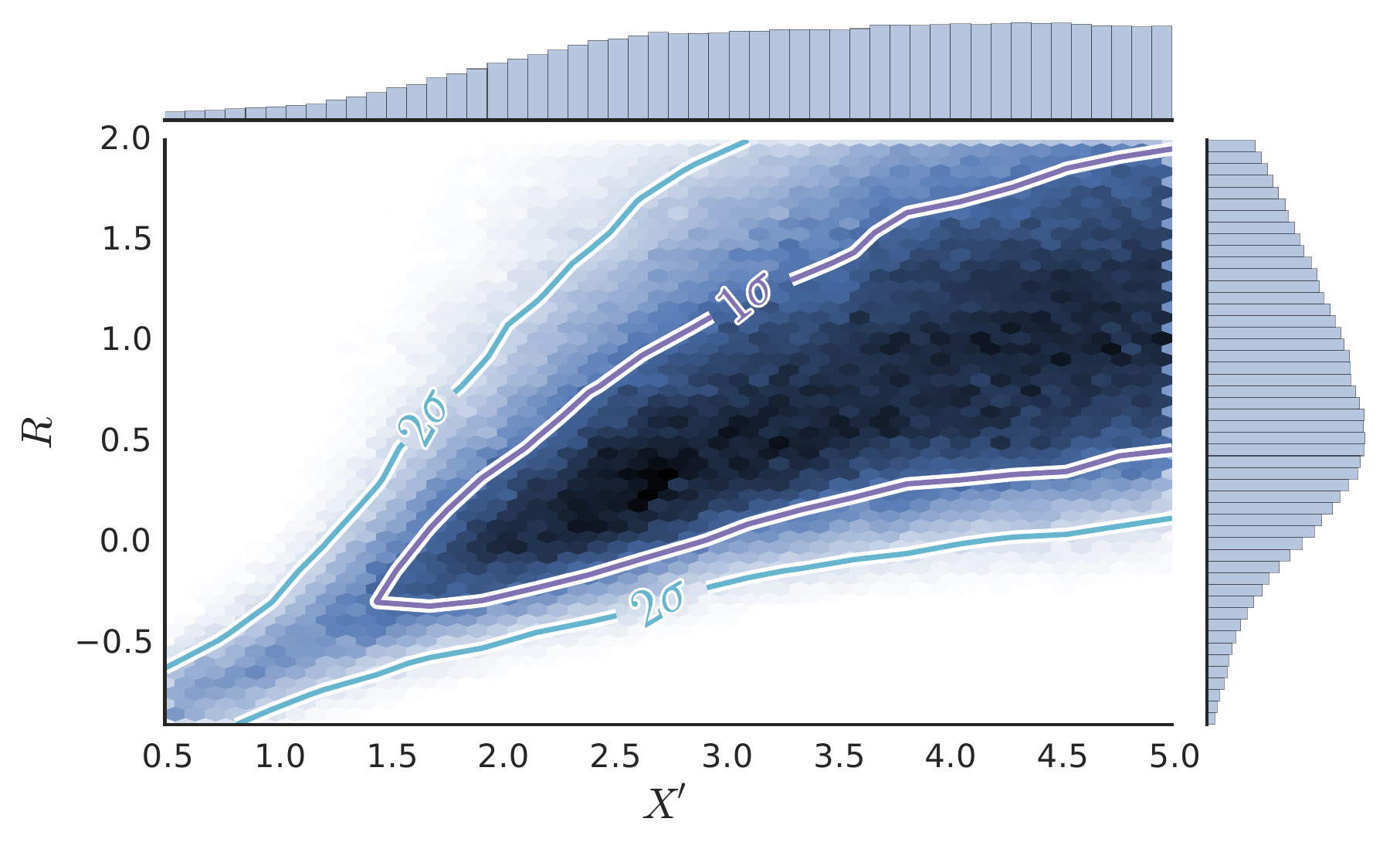}
\caption{\label{fig:posterior_eos}
The posterior likelihood for the two parameters that describe the equation of state, $X'$ and $R$, have a preference to be along the diagonal. This shows that experiment constrains some integrated measure of the overall stiffness of the equation of state, i.e. a softer equation of state just above $T_c$ is consistent with the data if it is combined with a more rapid stiffening at higher temperature.}
\end{figure}
\begin{figure}
\includegraphics[width=0.48\textwidth]{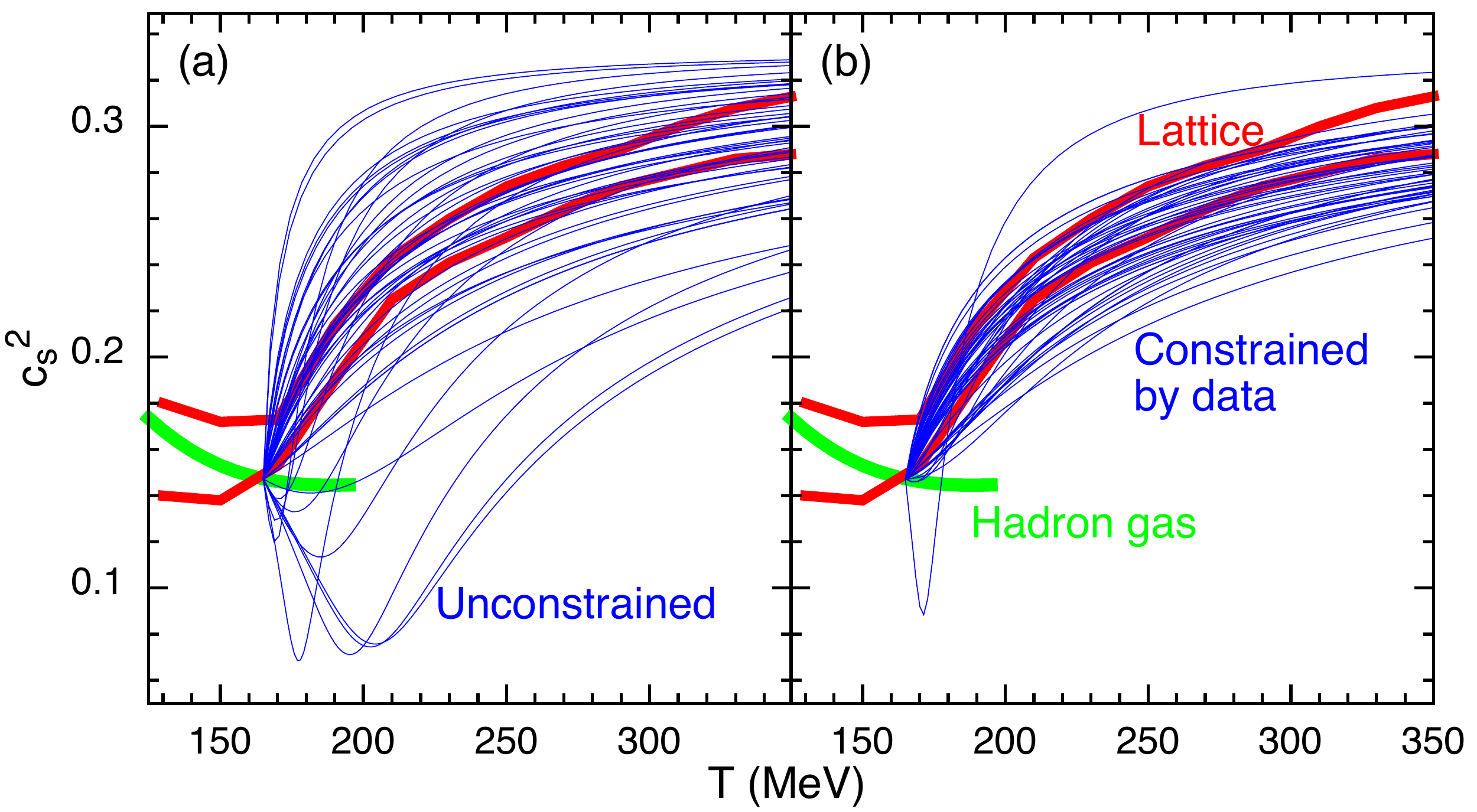}
\caption{\label{fig:priorvpost50}
(a) Fifty equations of state were generated by randomly choosing $X'$ and $R$ in Eq. (\ref{eq:EoSform}) from the prior distribution and weighted by the posterior likelihood (b). The two upper thick lines in each figure represent the range of lattice equations of state shown in \cite{Bazavov:2014pvz}, and the lower thick line shows the equation of state of a non-interacting hadron gas. This suggests that the matter created in heavy-ion collisions at RHIC and at the LHC has a pressure that is similar, or slightly softer, to that expected from equilibrated matter.}
\end{figure}

The ability of the procedure to constrain the two parameters that determine the equation of state is shown in Fig. \ref{fig:posterior_eos}. As a function of $X'$ and $R$ defined in Eq. (\ref{eq:EoSform}), the likelihood is significant for a large band near the diagonal. Higher values of $X'$, which delays the approach of the speed of sound to one third until higher energy densities and makes the equation of state softer, can be compensated by higher values of $R$, which sends the speed of sound higher just above $T_c$ and makes the equation of state stiffer. Fifty values of $X'$ and $R$ were then taken randomly from both the prior, and weighted by the posterior likelihood. For each case the speed of sound is plotted as a function of the temperature in Fig. \ref{eq:EoSform}. It is clear that the experimental results significantly constrain the equation of state and we also note that the RHIC and LHC data in combination provide a better constraint than either can alone. It appears that the speed of sound cannot fall much below the hadron gas value, $\sim 0.15$, for any extended range and that it must rise with temperature. Figure \ref{fig:priorvpost50} also shows a range of equations of state from lattice calculations \cite{Bazavov:2014pvz}. The equations of state found here show a preference for being slightly softer than those from the lattice, but the ranges overlap.

\section{Conclusions}

Determining the equation of state from experiment has proven difficult due to the intertwined links between model parameters and numerous observables. The statistical techniques applied here overcome these difficulties. The resulting constraints suggest the speed of sound gradually rises as a function of temperature from the hadron gas value. The band of equations of state from Fig. \ref{fig:priorvpost50} is modestly softer than that of lattice calculations, but has significant overlap. This analysis strengthens the supposition that the matter created in relativistic heavy ion collisions has properties similar to that of equilibrated matter according to lattice calculations and shows that our model describes the dynamics of heavy ion collisions well enough to permit the extraction the thermodynamic and transport properties of equilibrium condensed QCD matter.

\begin{acknowledgments}
This work was supported by the National Science Foundation's Cyber-Enabled Discovery and Innovation Program through grant NSF-0941373 and by the Department of Energy Office of Science through grant number DE-FG02-03ER41259. The authors thank Ron Soltz for providing the lattice data.
\end{acknowledgments}

\end{document}